\documentclass[mathleft,fleqn,%
]{an}
\usepackage{graphicx}
\usepackage[varg]{txfonts}
\usepackage{natbib}
\bibpunct{(}{)}{;}{a}{}{,}
  \def \teff {$T_{\mathrm{eff}}$}
  \def \tc {$T_{\mathrm{c}}$}
  
  \def \logg {$\log g$}

\begin{document}

\Pagespan{1}{}
\Yearpublication{2016}%
\Yearsubmission{2016}%
\Month{0}%
\Volume{}%
\Issue{}%
\DOI{}%

\title{Sun-like stars unlike the Sun \\
        Clues for chemical anomalies of cool stars}

\author{V. Adibekyan\inst{1}\fnmsep\thanks{Corresponding author:
        {vadibekayn@astro.up.pt}}
\and  E. Delgado-Mena\inst{1}
\and  S. Feltzing\inst{2}
\and  J.\,I. Gonz\'{a}lez Hern\'{a}ndez\inst{3,4}
\and  N.\,R. Hinkel\inst{5,6}
\and  A.\,J. Korn\inst{7}
\and  M. Asplund\inst{8}
\and  P.\,G. Beck\inst{9}
\and  M. Deal\inst{10,11}
\and  B. Gustafsson\inst{7,12}
\and  S. Honda\inst{13}
\and  K. Lind\inst{14,7}
\and  P.\,E. Nissen\inst{15}
\and  L. Spina\inst{16}
}
\titlerunning{Sun-like stars unlike the Sun}
\authorrunning{Adibekyan et al.}
\institute{
Instituto de Astrof\'isica e Ci\^encias do Espa\c{c}o, Universidade do Porto, CAUP, Rua das Estrelas, 4150-762 Porto, Portugal
\and 
Department of Astronomy and Theoretical Physics, Lund Observatory, Box 43, SE-22100 Lund, Sweden
\and
Instituto de Astrof\'{\i}sica de Canarias, 38200 La Laguna, Tenerife, Spain
\and 
Departamento de Astrof{\'\i}sica, Universidad de La Laguna, 38206 La Laguna, Tenerife, Spain
\and
School of Earth \& Space Exploration, Arizona State University, Tempe, AZ 85287, USA
\and
Vanderbilt University, Department of Physics and Astronomy, 6301 Stevenson Center Lane, Nashville, TN 37235, USA
\and
Department of Physics and Astronomy, Uppsala University, Box 516, SE-75120 Uppsala, Sweden
\and
The Australian National University, Research School of Astronomy and Astrophysics, Cotter Road, Weston, ACT, 2611, Australia
\and
Laboratoire AIM, CEA/DRF-CNRS-Univ. Paris Diderot-IRFU/SAp, Centre de Saclay, F-91191 Gif-sur-Yvette Cedex, France
\and
Laboratoire Univers et Particules de Montpellier (LUPM), UMR 5299, Universit\'{e} de Montpellier, CNRS, place Eug\`{e}ne Bataillon,
34095 Montpellier Cedex 5, France
\and
CNRS, IRAP, 14 avenue Edouard Belin, 31400 Toulouse, France
\and
NORDITA, Roslagstullsbacken 23, SE-106 91 Stockholm, Sweden
\and
Nishi-Harima Astronomical Observatory, Center for Astronomy, University of Hyogo, 407-2 Nishigaichi, Sayo-cho, Sayo, Hyogo 679-5313, Japan
\and
Max-Planck Institut f\"{u}r Astronomie, K\"{o}nigstuhl 17, 69117 Heidelberg, Germany
\and 
Stellar Astrophysics Centre, Department of Physics and Astronomy, Aarhus University, Ny Munkegade 120, 8000, Aarhus C, Denmark
\and
Universidade de S\~{a}o Paulo, Departamento de Astronomia do IAG/USP, Rua do M\~{a}tao 1226, S\~{a}o Paulo, 05509-900, SP, Brasil}

\received{XXXX}
\accepted{XXXX}
\publonline{XXXX}


\abstract{%
  We present a summary of the splinter session  ``\textit{Sun-like stars unlike the Sun}`` that was held on 09 June 2016 as
part of the Cool Stars 19 conference (Uppsala, Sweden). We discussed the main limitations (in the theory  and observations) in the derivation of very precise stellar parameters 
and chemical abundances of Sun-like stars. We outlined and discussed the most important and most debated processes 
that can produce chemical peculiarities in solar-type stars. Finally, in an open discussion between all the participants we tried to identify new pathways and 
prospects towards future solutions of the currently open questions.
}

\maketitle

\section{Motivation}

In stellar astronomy, we sometimes divide stars into two wide groups and colloquially refer to them as cool stars (late-type stars) 
and hot stars (early-type stars), although there is no sharp division between these two groups. 
These kinds of definitions are lexical and have only a descriptive, qualitative character. A more quantitative boundary between
cool and hot stars was suggested by \citet{Gray-05} based on the shape of the bisectors of the spectral lines. Main sequence stars 
with spectral types of later than about F0 have a bisectors with a so-called classical C shape, while hotter stars show reversed C shape.
This boundary is called ''granulation boundary`` \citep[e.g.][]{Gray-89, Gray-86} and practically divides the 
Hertzsprung--Russell diagram into cool and hot stars \citep{Gray-05}. 
In this work, adopting the definition of Gray we refer to stars later than F0 when saying ''cool stars``. These low-mass stars have long lifetimes and their 
envelopes contain information about their stellar evolution and the history of the evolution of chemical abundances in the Galaxy.

During the last decade, noticeable advances were made in the characterization of atmospheric properties 
(e.g. effective temperature, metallicity, surface gravity) and chemical abundances of cool stars. 
The high precision in stellar atmospheric parameters is crucial for precise characterization 
of physical properties of stars such as their mass and age.  

The extremely high precision in chemical abundance derivations allowed observers to study subtle 
chemical peculiarities in Sun-like stars. Given the nature of the 
detailed chemical abundance derivations, it is likely that many physical processes determine the 
chemical characteristics of the stars. 
Understanding the origin of these anomalies is very 
important for the further advancement of Galactic and stellar astronomy, as well as the very fast advancing 
field of exoplanetary research.

In June 2016, we organized a Splinter Session at the Cool Stars 19 workshop with the goal to bring together experts of stellar, Galactic and 
planetary astrophysics to highlight the latest results and discuss what may make Sun-like stars unlike the Sun. We had six invited review talks and 
four contributed talks, which were followed by an open discussion between speakers and participants. The main scientific questions 
discussed during the session were:

I. Abundances of the Sun and Sun-like stars. 
What is the highest precision and accuracy we actually can expect current analysis methods to deliver? 
What are the limitations in the theory (e.g. model of atmospheres, 1D, hydrostatic, LTE) and observations (e.g. spectral resolution, 
signal-to-noise ratio, atmospheric observational conditions)? 

II. Abundance characteristics of stars. 
How is the inhomogeneous Galactic chemical evolution, the star- and planetary formation history, 
and the stellar evolution reflected in the surface abundances of Sun-like stars?
How can we study these different aspects by analyzing the elemental abundances in stellar spectra?

\noindent

In this paper we summarize the presentations and the discussions of this splinter session.

\section{Solar twins, analogs and solar-type stars}

Classifying a star as solar-type, solar analog, or solar twin depends on the degree of similarity between the star and the Sun. 
The categorization also reflects the evolution of astronomical instrumentation and observational techniques.
\citet{CayreldeStrobel-96} defined a solar twin as a star that has the same atmospheric and physical properties 
as the Sun within the observational errors. This definition obviously depends on the uncertainties of the derived parameters.
\citet{Soderblom-98} provided a more practical definition of these three categories of stars.
While the literature is full of quantitatively different definitions of solar twins, analogs and sun-like (solar type) stars, these definitions 
are qualitatively similar \citep[e.g.][]{Melendez-09, Ramirez-09, GH-10, Adibekyan-14, doNascimento-14, PortodeMello-14, Datson-15}. 
For more discussion on different definitions we refer the reader to \citet{Datson-14}.

In the splinter session the following definitions for solar twins, analogs and Sun-like stars in terms of stellar parameters were presented and used. Solar
twins: \teff \ = 5777\footnote{We note that the value of the nominal solar effective temperature recomended by IAU 2015 resolution B3
is 5772$\pm$0.8K \citep{Prsa-16}.}$\pm$100 K, \logg \ = 4.44$\pm$0.10 dex, [Fe/H] = 0.00$\pm$0.10 dex \citep[e.g.][]{Ramirez-09, Adibekyan-14},  solar analogs: 
\teff \ = 5777$\pm$200 K, \logg \ = 4.44$\pm$0.20 dex, [Fe/H] = 0.00$\pm$0.20 dex \citep[e.g.][]{Adibekyan-14}, and 
solar-type:  main sequence or subgiant stars with 5000 K $<$ \teff \ $<$ 6500 K.

With the recent advances of asteroseismology, thanks to \textit{Kepler} \citep{Borucki-10}   and \textsc{CoRoT}  
\citep[Convection, Rotation, and planetary Transits --][]{Baglin-06}  missions, parameters determined by astroseismology have also been  included in the
definition of solar analogues and twins. In particular, the presence of solar-like oscillations can be used to consider a star as a solar analog, or a
seismic solar analog \citep[e.g.][]{Metcalfe-12, doNascimento-13, Salabert-16a, Beck-16}.

\subsection{Accuracy and precision in stellar parameters and chemical abundances}

High-precision and high-accuracy stellar abundances are crucial for many fields of stellar, planetary and galactic astrophysics. 
However, precise and accurate derivation of stellar atmospheric abundances is a difficult challenge which is obvious when comparing different 
techniques and measurements \citep{Hinkel-16, Jofre-17}.

\subsubsection{High-resolution spectroscopy}

If past analyses of large, homogeneous and high-quality data  reached abundance precisions of 0.03-0.07 dex 
\citep[e.g.][]{Bensby-03, Valenti-05, Reddy-06, Gilli-06, Takeda-07, Nissen-10, Adibekyan-12, Adibekyan-15, Bensby-14},
the latest works on solar twins that are based on differential line-by-line analysis report even higher precision of $\lesssim$0.01 dex
\citep[e.g.][]{Melendez-09, GH-10, GH-13, Ramirez-10, Melendez-10, Adibekyan-16a, Adibekyan-16b, TucciMaia-14, Bedell-14, Nissen-15, Nissen-16, Saffe-16, Spina-16a, Spina-16b}.
Consequently, the precision in atmospheric parameters reported for the solar twins is very high: $\sim$10 K for \teff, $\sim$0.02 dex for \logg,
and $\sim$0.01 dex for [Fe/H] \citep[e.g.][]{Ramirez-14, TucciMaia-14, Bedell-14, Adibekyan-16a, Spina-16a, Spina-16b}. 

\begin{figure}
\includegraphics[width=\linewidth]{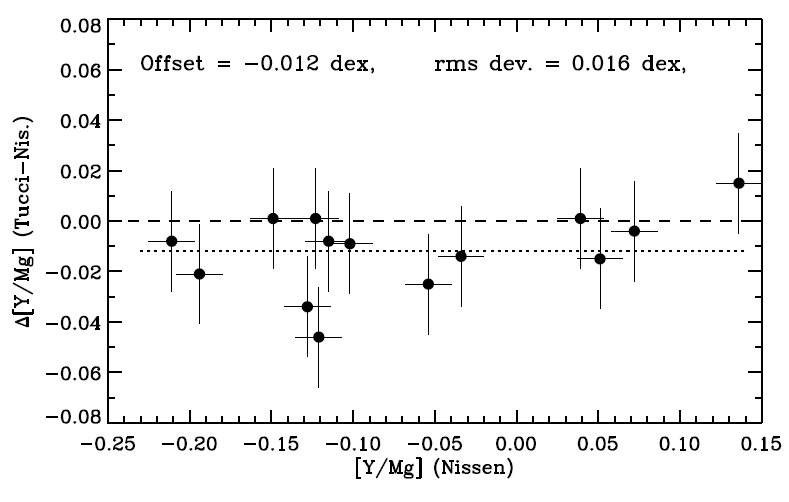}
\caption{Comparision of [Y/Mg] abundance ratios derived by \citet{Nissen-16} and \citet{TucciMaia-16} for 14 solar twin stars. 
(Courtesy of Poul Erik Nissen).}
\label{figure_nissen_tucci}
\end{figure}

Recently, \citet{Bedell-14} analyzed  solar  spectra  observed with different instruments, from different asteroids, and
at different times, i.e, conditions. The authors reached a conclusion that a major effect on differential relative abundances is
caused by the use of different instruments (up to 0.04 dex). They found that the choice of asteroids to obtain 
the solar reflected spectra and time-dependent effects (observations at different epochs) are smaller than 0.01 dex.
\citet{Bensby-14} have also analyzed (applying exactly the same analysis techniques) different spectra of the Sun 
(scattered solar light from the afternoon sky, the Moon, Jupiter's moon Ganymede, and the asteroids
Vesta and Ceres) obtained during a period of six years with different instruments. The maximum observed differences for solar parameters 
were 49 K for \teff, 0.03 dex for \logg, and 0.03 dex for solar metallicity.
In turn, \citet{Adibekyan-16a} showed that the average difference in chemical abundances observed for two different 
high-quality (signal-to-noise ratio of about 400) spectra obtained during the same night for the same star is usually small
$\lesssim$0.01$\pm$0.03 dex, but can reach up to 0.06 dex depending on the element.
 
Most of the reported uncertainties are in fact precision, or internal (random) errors (e.g., uncertainties in the continuum setting, in the $\log gf$ values). Systematic errors, due to the 
model atmospheres and atomic data are more difficult to estimate and can be much larger than the random errors. Recently, \citet{Bensby-14}  evaluated
the external precision of their abundance derivation by comparing their results for solar-type stars with those from the literature \citep{Reddy-03, Reddy-06, Valenti-05, Adibekyan-12}.
The authors found that the differences (average for all stars in common) in stellar parameters and abundances of individual elements observed 
between different works range from -10K to +120 K for \teff,
from -0.05 to -0.07 dex for \logg, from -0.02 to +0.03 dex for [Fe/H], and from -0.09 to 0.10 dex for different elements. For more complete and extensive comparison of
more than 80 data sets we refer the reader to \citet{Hinkel-14}. They found that the variation between studies per element has a mean of 0.14 dex 
for all elements in all stars in their compiled catalog, called the Hypatia Catalog.

The comparison between different studies of solar twin stars shows higher agreement. In particular, \citet{Nissen-15} when comparing his results 
with those of \citet{Ramirez-14} for 14 stars in common obtained an average difference and $rms$ deviation of: $\Delta$\teff \ = 0$\pm$10 K, 
$\Delta$\logg \ = 0.002$\pm$0.020 dex, and  $\Delta$[Fe/H] = 0.000$\pm$0.014 dex. The same author, when comparing his results with that of \citet{Sousa-08} 
for the 21 solar twins in common found the following average differences and rms deviations: $\Delta$\teff \ = -1$\pm$8 K, 
$\Delta$\logg \ = 0.018$\pm$0.033 dex, and  $\Delta$[Fe/H] = -0.003$\pm$0.009 dex. Comparison of chemical abundances of individual elements, and abundance ratios
is also usually small. For example, the average offset and $rms$ deviation in [Y/Mg] abundance ratio observed between \citet{Nissen-16} and \citet{TucciMaia-16}
is 0.012$\pm$0.016 dex (see Fig.~\ref{figure_nissen_tucci}).

\subsubsection{3D and non-LTE effects}

Most of the studies, when deriving stellar parameters and elemental abundances used classical 1D hydrostatic models with an assumption of local 
thermodynamic equilibrium (LTE). Thanks to the exponentially increasing level of computational power, huge progress has been made in the last decade 
in developing 3D hydrodynamical model atmospheres \citep[e.g.][]{Magic-13, Freytag-12, Beeck-13, Trampedach-13}. Furthermore, non-LTE calculations and corrections 
are now available for more than 20 elements e.g. Li, O, Na, Mg, Si, Ca, Ti, Fe, Sr, Ba \citep[e.g.][]{Lind-09, Amarsi-15, Osorio-15, Bergemann-11, Lind-12, Korotin-15,
Merle-11, Shi-11, Spite-12, Prakapavicius-13, Amarsi-16}. For a detailed discussion of non-LTE effects in the lines of different elements we refer the reader to 
\citet{Mashonkina-14}.

\begin{figure*}
\includegraphics[width=0.9\linewidth]{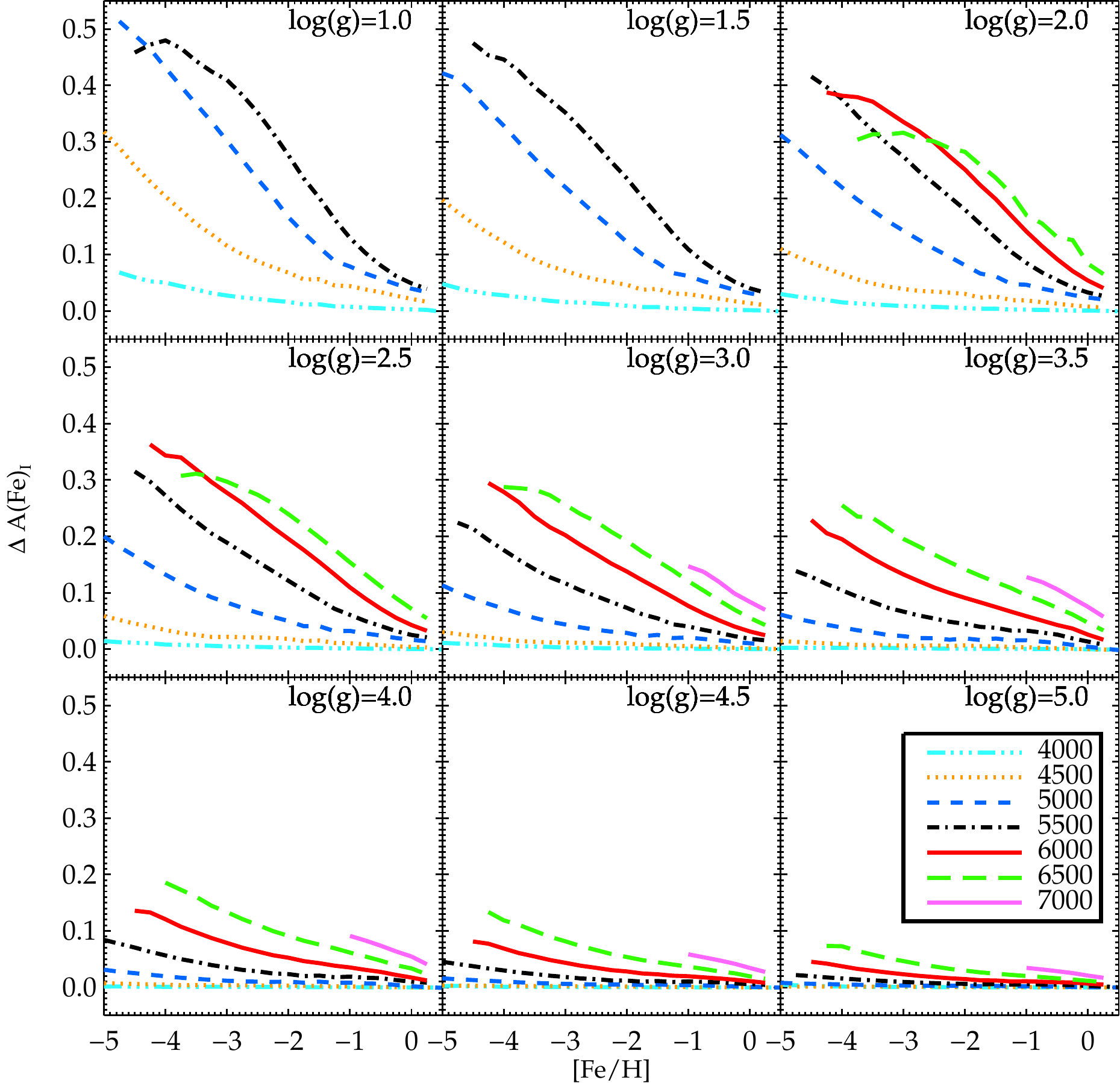}
\caption{The dependence of typical NLTE corrections for high-excitation (E$_{exc} >$ 2.5 eV), unsaturated ($W_{\lambda} <$ 50 m$\AA{}$) Fe I lines on
stellar parameters. All models have $\xi_{t}$ = 2.0 km s$^{−1}$. The figure is from \citet{Lind-12}.}
\label{figure_fe_nlte}
\end{figure*}

For most elements with complex atoms, non-LTE effects are not very strong in solar-twins. The amplitude of these effects are different for different 
species (species sensitive to over-ionisation or collision-dominated species) and depends on atmospheric parameters of the stars. For example in the case of 
iron (see Fig.~\ref{figure_fe_nlte}) and other neutral Fe-peak atoms, the non-LTE effects 
increase as temperature increases,  surface gravity decreases, and the metallicity decreases \citep[e.g.][]{Bergemann-14}. 
Obviously, when comparing stars with very similar parameters, such as solar twins, the differential non-LTE effects are very small \citep[e.g.][]{Nissen-15, Spina-16a}.
They become non-negligible for high-precision work on solar analogs and should be considered when  solar-type stars are intercompared.

\subsubsection{Asteroseismology}

Comparison of physical parameters derived with different, independent  methods help us to understand and estimate the accuracy of the derivations. A combination
of different astronomical tools and methods also helps to improve the accuracy of the determinations. In particular, asteroseismology combined with 
high-resolution spectroscopy allows us to substantially improve the accuracy of the stellar parameters \citep[e.g.][]{Chaplin-14, Metcalfe-14, Lebreton-14, Creevey-17}. 
The role of asteroseismology is particularly invaluable for derivation of stellar ages, which is usually 
very difficult to determine with high accuracy using spectroscopy. Asteroseismology can provide ages of solar-like stars with a relative precision of
10 to 15 \% \citep[e.g.][]{Chaplin-14}.

Such improved stellar parameters are beneficial to study aspects of stellar structure and evolution such as rotation, 
activity or the lithium abundance. Examples of such an effort are the works of \citet{Salabert-16a} and \citet{Beck-17}
where the authors carefully selected 18 seismic solar analogs to study different properties of these stars\footnote{
A summary of the current work on this sample of 18 solar analogues is described in dedicated proceeding articles by \citet{Beck-16a} and \citet{Salabert-16b}.}.
\citet{Beck-17} used the Toulouse-Geneva Evolutionary Code \citep[TGEC,][]{Hui-Bon-Hoa-08},
to compute the theoretical evolution of lithium for the asteroseismic mass. 
TGEC includes complete atomic diffusion (including radiative accelerations) and non-standard mixing processes.
Then the authors compared it to the spectroscopically 
measured lithium abundance and seismically determined age. For all the stars they found a good agreement within the realistic uncertainties of stellar mass and age 
derived by \citet{Lebreton-14}. Such agreement between theory and observations suggests that 
for these solar analogues, the same physical processes are driving internal mixing.

\section{Galactic chemical evolution and nucleosynthesis with solar twins}

Important information about the formation and evolution of galaxies are locked into the chemical compositions of stars. 
All the metals, or elements heavier than Boron, originate from stars that enrich the interstellar medium with their own unique pattern of 
elements depending on their mass and initial metallicity. In fact, each specific element has been produced by 
different sites of nucleosynthesis that contribute to the chemical evolution of galaxies with different timescales \citep[e.g.][]{Pagel-09}. 

The chemical abundances, measured as [X/Fe] vs. [Fe/H] for solar-type stars are traditionally used to study the Galactic 
chemical evolution because iron has been assumed to be a good chronological indicator of nucleosynthesis \citep[e.g.][]{Bensby-03, Adibekyan-12,
Smiljanic-16, Romano-10, Edvardsson-93, Chiappini-97, Smiljanic-14}. Obviously, the studies of the 
relations between the  abundance ratios and age would provide more direct information about the 
nucleosynthetic history of elements and chemical evolution of our Galaxy.

\begin{figure}
\includegraphics[width=\linewidth]{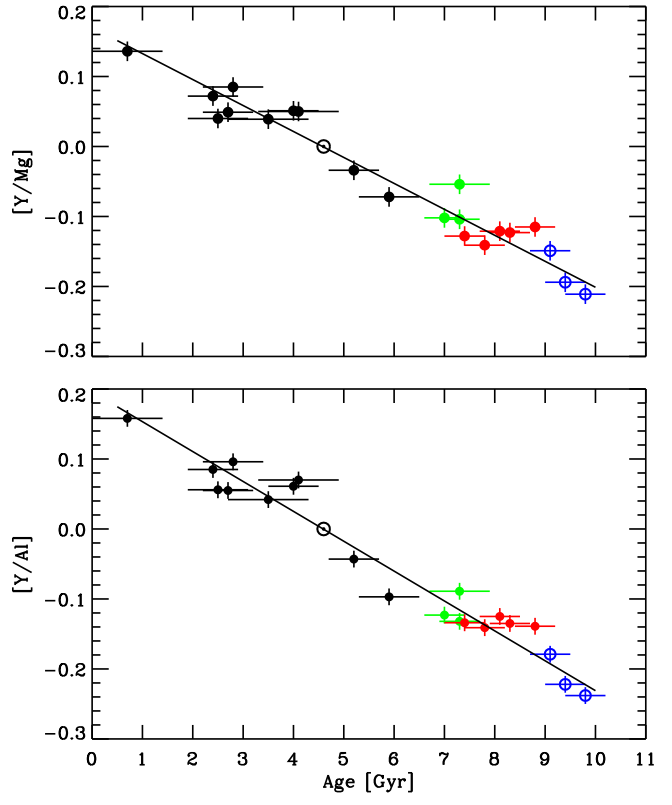}
\caption{[Y/Mg] and [Y/Al] versus stellar age. Stars younger than 6 Gyr are marked with black filled circles.
 Old stars with low [Na/Fe] and high [Na/Fe] are shown in red and green filled circles, respectively. 
Three [$\alpha$/Fe]-enhanced stars are shown with open blue circles and the Sun is shown with its typical $\odot$ symbol. 
The figure is from \citet{Nissen-16}.}
\label{figure_nissen_y_mg}
\end{figure}

Recently, \citet{Nissen-15, Nissen-16}  used relatively high-precision ages (derived from evolutionary tracks) and  chemical abundances of 18 elements (from O to Ba)
determined for 21 solar twins to study the correlations between these two parameters. For stars younger than 6 Gyr, Nissen found 
that some elements show very tight correlation with stellar age. Nissen showed that this linear correlation breaks down at 6 Gyr
and the stars with ages between 6 and 9 Gyr split up into two groups with high and low values of [X/Fe] for the 
odd-Z elements  Na,  Al,  Sc,  and Cu. \citet{Nissen-16} concluded that the younger stars were formed from a well-mixed interstellar 
gas while older  stars formed in regions  that were enriched by supernovae with different neutron  excesses.
He also showed that due to very tight linear correlation with age, [Y/Mg] and  [Y/Al]  abundance ratios
can be used to derive stellar ages with a precision reaching 1 Gyr (see Fig.~\ref{figure_nissen_y_mg}). This result on solar twins was later confirmed by other authors
\citep{Spina-16a, TucciMaia-16} and was extended to solar analogs \citep{Adibekyan-16b}. Interestingly, \citet{Feltzing-16} recently showed 
that the correlation between [Y/Mg] and age is a function of metallicity and gets flat at metallicities below -0.5 dex.

\begin{figure}
\includegraphics[width=\linewidth]{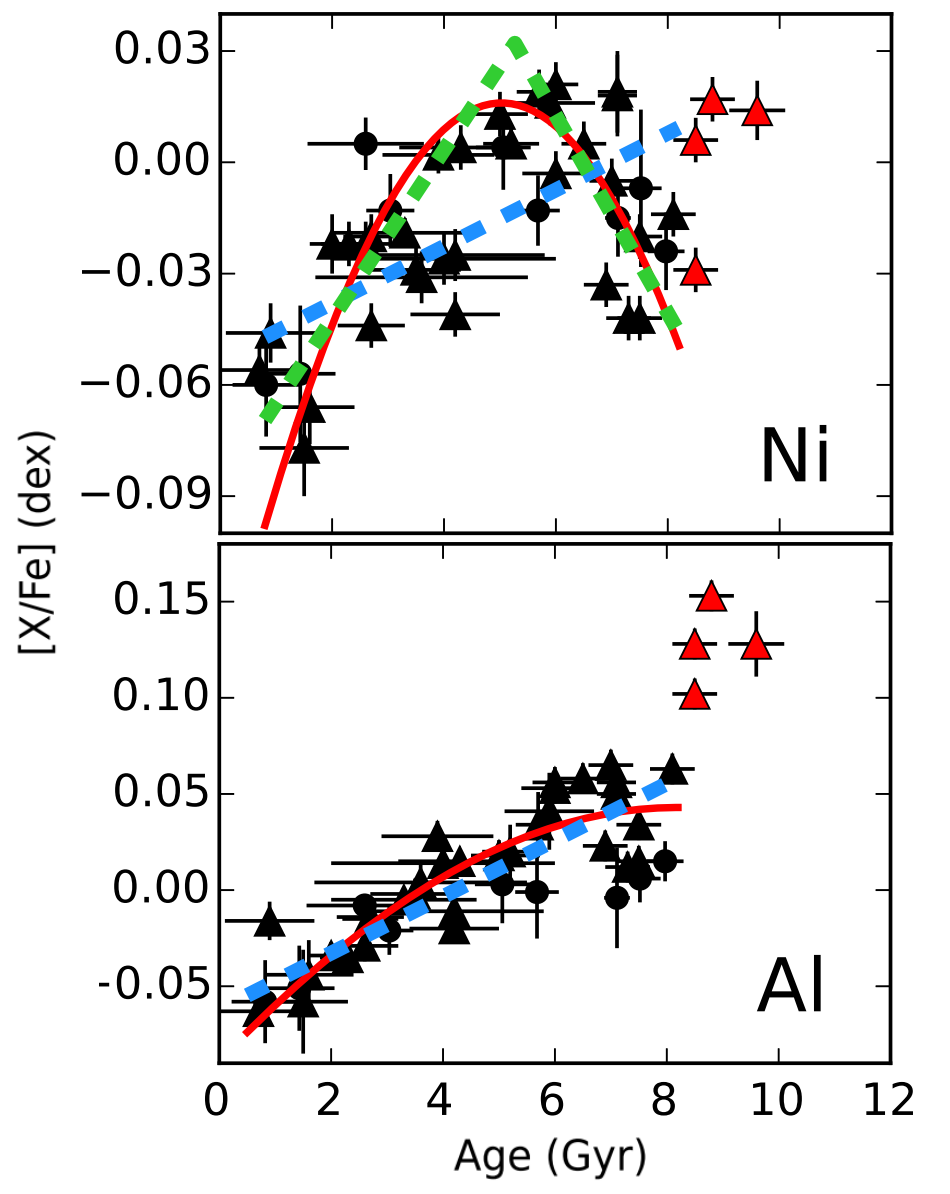}
\caption{[Al/Fe] and [Ni/Fe] ratios as a function of stellar ages. The red triangles represent the thick disk stars, while the black symbols are the thin disk stars. 
The [X/Fe]-age relations have been fitted by linear, hyperbolic and two-segmented line functions shown as blue dashed, red solid and green dashed lines, 
respectively. (Courtesy of Lorenzo Spina).}
\label{figure_ni_al}
\end{figure}

More recently, \citet{Spina-16a} studied a sample of 41 thin disk and four thick disk stars for which superb abundances with 0.01 dex 
precision and accurate stellar ages have been obtained through a line-by-line differential analysis of the 
EWs relative to the solar spectrum  \citep[see][]{Bedell-14, Spina-16b}.
Based on this data set, \citet{Spina-16a} outlined the [X/Fe]-age relations over a time interval of 10 Gyr (see Fig.~\ref{figure_ni_al}). They presented the 
[X/Fe] - age relations for 23 elements (C, O, Na, Mg, Al, Si, S, Ca, Sc, Ti, V, Cr, Mn, Co, Ni, Cu, Zn, Y, Ba, La, Ce, Nd, and Eu).
Their main results revealed that each different class of elements showed distinct evolution with time that relies on the different characteristics, 
rates and timescales of the nucleosynthesis sites from which they are produced. The $\alpha$-elements are characterized by a 
[X/Fe] decrement as time goes on. Strikingly, an opposite behavior is observed for Ca. The iron-peak elements show an early 
[X/Fe] increase followed by a decrease towards the youngest stars. The [X/Fe] for the n-capture elements decrease with age.

Knowledge of the [X/Fe]-age relations is a gold mine from which we can achieve considerable understanding about the processes 
that governed the formation and evolution of the Milky Way: the nature of the star formation history, the supernovae (SNe) rates, 
the stellar yields, and the variety of the SNe progenitors, etc. This approach has been already successfully applied before to low(er) 
precision data \citep[e.g.][]{Edvardsson-93} and demonstrated its power. These types of studies are of fundamental significance 
in efforts to reconstruct the nucleosynthethis history of the Galactic disk through chemical evolution models.

\section{Chemical abundances of Sun-like stars with and without planets}

The connection (probably bi-directional) between stellar and planetary properties has been widely explored.
In particular, the very first correlation observed in the field of exoplanetary research was the correlation between the giant-planet occurence and stellar metallicity 
(usually iron content was used as a proxy for overall metallicity) \citep[e.g.][]{Gonzalez-97, Santos-01, Santos-04, Fischer-05, Sousa-11}. 
Later, studies based on large and homogeneous data-sets showed that elements other than iron, such as, C, O, Mg, and Si, may play a very 
important role for planet formation \citep[e.g.][]{Robinson-06, Haywood-09, Adibekyan-15a, Adibekyan-12a, Adibekyan-12b, Brugamyer-11, DelgadoMena-10}.

Interestingly, the importance of stellar (and disk) metallicity is likely not only limited to the formation of planets.
It is now becoming clear that the architecture, structure and even habitability of planets strongly
depend on the chemical properties of their hosts \citep[e.g.][]{Dawson-13, Adibekyan-13, Adibekyan-16c, Santos-15, Dorn-15}. 
In particular we see that the position of planets in the period-mass diagram depends on the metallicity of the host star \citep[][]{Beauge-13, Adibekyan-13}.
We learned that the presence or absence of gaseous atmosphere of small-sized planets probably depends on the metallicity \citep[][]{Dawson-15}and
we know that mineralogical ratios, such as Mg/Si and Fe/Si, may control the structure and composition of terrestrial planets 
\citep[e.g.][]{Grasset-09, Bond-10, Thiabaud-14, Dorn-15}. These results imply that the study of the link between planet
formation processes and properties of their hosts will lay the groundwork to answer many questions related to  formation and evolution of both planets and stars.

\subsection{Tc trend}

After the first planets were discovered, astronomers tried to search for chemical signatures of planet
formation and planet engulfment on the planet-host stars. Several studies, starting from \citet{Gonzalez-97} and \citet{Smith-01}, explored a possible 
trend between the abundances
of chemical elements and the condensation temperature (\tc) of the elements. This trend is usually called ''\tc \ trend``, and the slope of the
correlation (slope of the linear fit) of [X/Fe] vs. condensation temperature is usually named ''\tc \ slope``. 

\begin{figure}
\includegraphics[width=\linewidth]{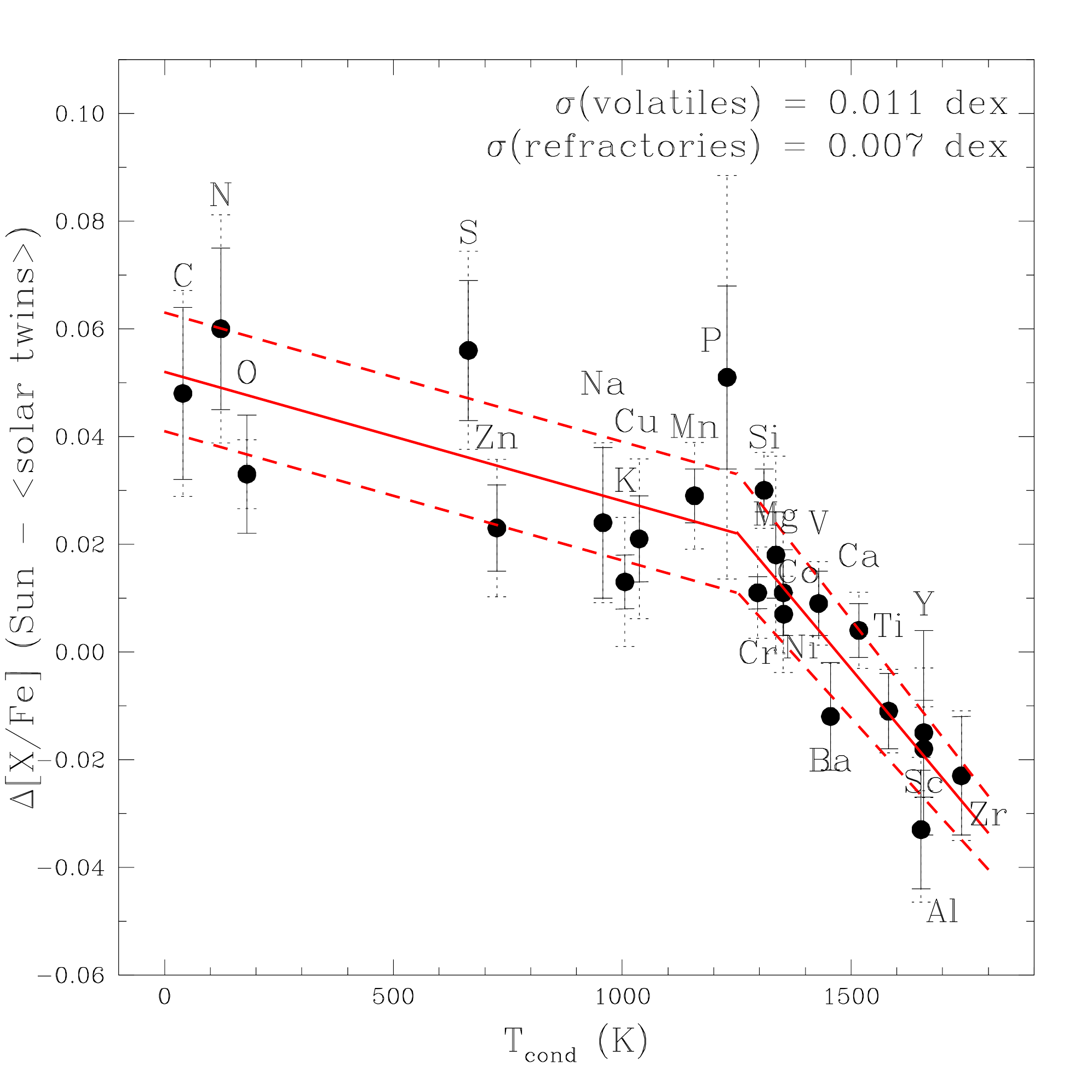}
\caption{Differences between [X/Fe] of the Sun and the mean values in the
solar twins (with no detected planets) as a function of T$_{cond}$. The abundance pattern shows a break at
T$_{cond}$ $\sim$ 1200 K. The solid lines are fits to the abundance pattern, while the
dashed lines represent the standard deviation from the fits.   The figure is from \citet{Melendez-09}.}
\label{figure_tc_melendez}
\end{figure}

\begin{figure*}
\includegraphics[width=1.0\linewidth]{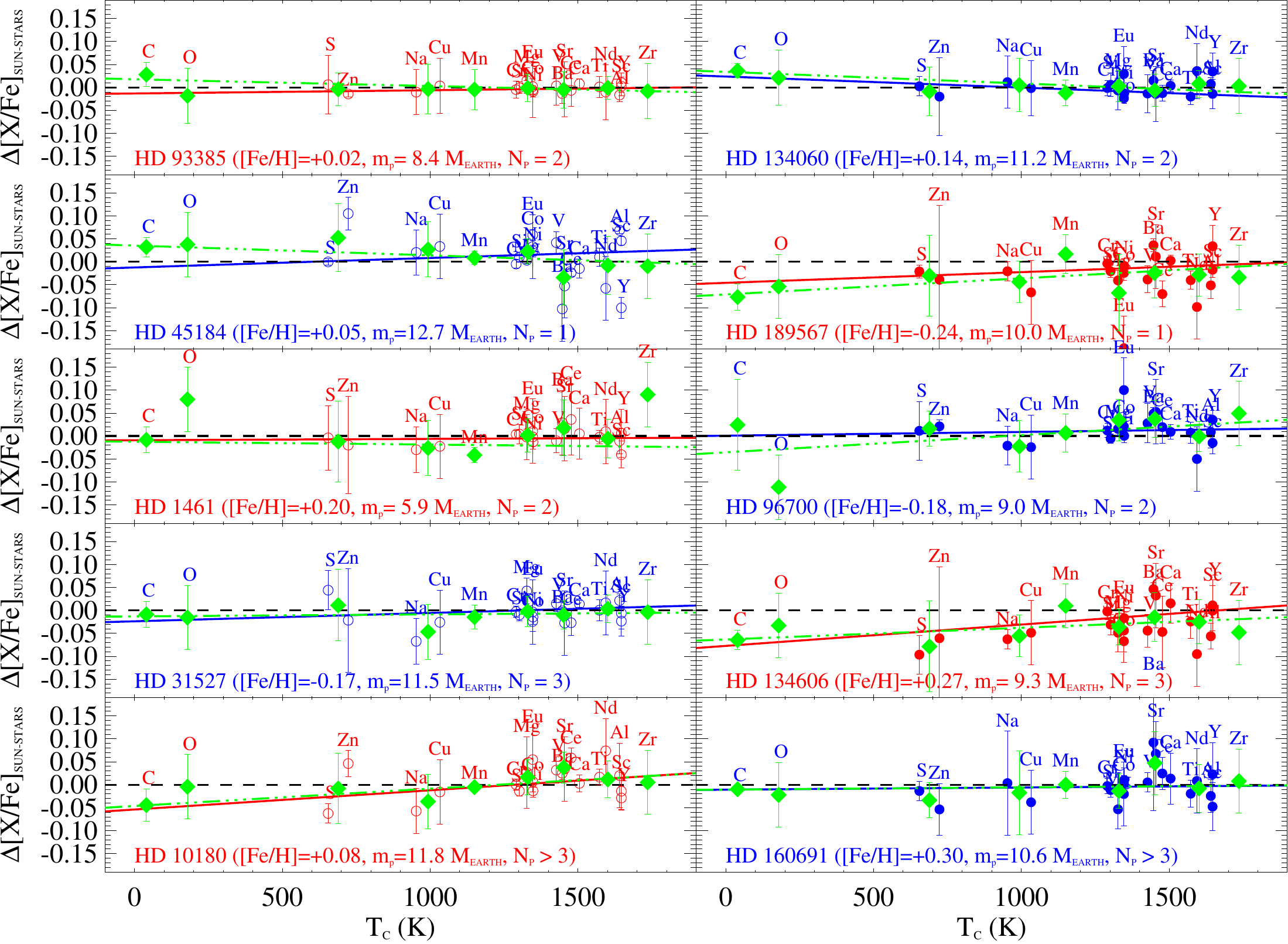}
\caption{Abundance differences, $\Delta$[X/Fe]$_{SUN-STARS}$, between the Sun and 10 stars hosting super-Earth-like planets (circles).
Diamonds show the average abundances in bins of $\Delta$\tc \ = 150 K. 
Linear fits to the data points (solid line) and to the mean data points (dashed-dotted line) weighted with the error bars are also displayed. 
The figure is from \citet{GH-13}.}
\label{figure_tc_gh13}
\end{figure*}

\citet{Melendez-09} were the first to report a statistically significant deficit of refractory elements (high-\tc) with respect to volatiles (low-\tc) in the Sun 
compared to solar twin stars (see Fig.~\ref{figure_tc_melendez}). The authors suggested that these missing elements were trapped in the terrestrial 
planets in our solar system. The same conclusion was also reached by \citet{Ramirez-09}, who analyzed a larger number of solar twins and analogs with and 
without detected planets. However, these results and explanations were strongly contested by \citet{GH-10} and \citet{GH-13}, who did not find a statistically
significant and consistent \tc \ trend when comparing stars with and without planets, even when evaluating these \tc \ trends for stars with detected super-Earth like planets
(see Fig.~\ref{figure_tc_gh13}).  This very exciting possible  connection between chemical peculiarities of parent stars and formation of
planets has also been examined in other works \citep[e.g.][]{Takeda-01, Ecuvillon-06, Sozzetti-06, 
Schuler-11,  Maldonado-15, Nissen-15, Biazzo-15, Saffe-15, Saffe-16, Mishenina-16, Hinkel-13, Spina-16a, Spina-16b}, but contradictory conclusions were reached.

\begin{figure}
\includegraphics[width=\linewidth]{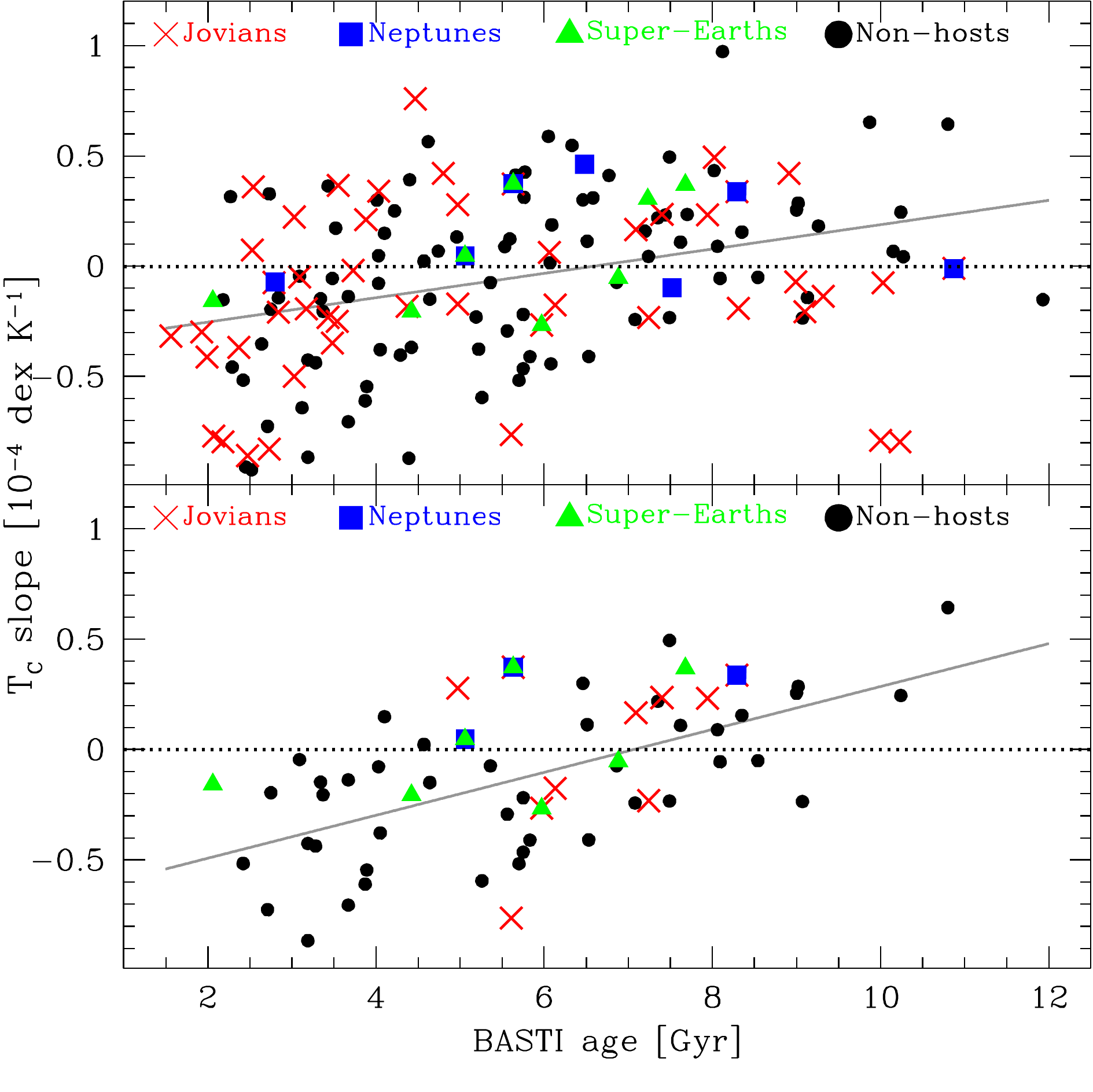}
\caption{\tc \ slopes versus ages for the full sample (top) and for the solar
analogs (bottom). Gray solid lines provide linear fits to the data points. The figure is from \citet{Adibekyan-14}.}
\label{figure_tc_age}
\end{figure}

Together with the rocky material accretion \citep[e.g.][]{Schuler-11, Spina-15} and/or rocky material trap \citep[e.g.][]{Melendez-09} in terrestrial planets, several explanations are proposed to explain the \tc \ trend.
\citet{Adibekyan-14} suggested that the \tc \ trend strongly depends on the stellar age (see Fig.~\ref{figure_tc_age}) and they found
a tentative dependence on the galactocentric distances of the stars. The correlation with stellar age was later confirmed by several authors 
\citep[e.g.][]{Nissen-15, Spina-16b}, while 
the possible relation with the galactocentric distances is more challenging \citep[see][]{Adibekyan-16b, Maldonado-15} probably because of its very complex nature
or because the galactocentric distances were estimated indirectly. 
\citet{Maldonado-15} and \citet{Maldonado-16}
further suggested a significant correlation with the stellar radius and mass. \citet{Onehag-14} in turn showed that while
the Sun shows a different \tc \ trend when compared to the solar-field twins, it shows a very similar abundance trend with \tc \ when
compared to the stars from the open cluster M67. They suggested that the Sun, unlike most stars, was formed in a dense
stellar environment where the protostellar disk was already depleted in refractory elements by radiative pressure on dust grains from bright stars before the Sun formed
\citep[see][for further discussion]{Gustafsson-16}. Gustafsson, at this meeting and in a forthcoming paper, has demonstrated that 
it is difficult in this way to cleanse enough material for forming a full cluster with such abundance characteristics -- the photoionization of 
the gas limits the amount of cleansed gas that is cool enough for star formation severely. 
\citet{Gaidos-15}  also suggested that gas-dust segregation in the disk can produce the \tc \ trend, although only a qualitative analysis and discussion was made.

To separate the possible chemical signatures of planet formation from the effects of Galactic chemical evolution, several authors
tried to correct the \tc \ slope by using the [X/Fe]--age relation \citep[e.g.][]{Yana-16, Spina-16b}. However, such kind of corrections are not easy to perform
because of the intrinsic scatter in the [X/Fe]--age distributions due, for instance, to migration processes in the Galaxy 
\citep[e.g.][]{Sellwood-02, Haywood-08, Haywood-13, Minchev-13} and possible intercorrelation between different parameters.

However, the comparison of binary systems of twin stars should not be affected by the above mentioned processes and effects (e.g. formation time and place)
and the only complications can be related to stellar evolution (if the stars do not have exactly the same physical properties e.g. mass).
Several authors studied the \tc \ trend in binary stars with and without planetary companions \citep[e.g.][]{Liu-14, Saffe-15, Mack-16}
or in binary stars where both components host planets \citep[e.g.][]{Biazzo-15, Teske-15, Ramirez-15, Teske-16}. 
Although some significant differences between the twin pairs
in some systems were reported, in general the results and conclusions of these studies point in different directions. 
Thus, as a whole, it is difficult to conclude that there are systematic differences in the chemical abundances of stars with and without planets in the binary systems. 
Moreover, there are discrepancies in the results even for the same individual systems such as 16 Cyg AB \citep[e.g. ][]{Laws-01, Takeda-05, Schuler-11a, TucciMaia-14}.
It should be noted also that there are not many high-precision abundance studies of binary stars where none of the stars host 
planets\footnote{One should always bear in mind that detection of low-mass/small-sized planets, especially at large separations, is very hard and 
the presence of this undetected planets is always possible and very probable since these planets are very common \citep[e.g.][]{Mayor-14, Mulders-16}.}. 
These kind of studies might help us to understand what is the largest detectable chemical anomaly not related to terrestrial planet formation.

\subsection{Li abundance}

Lithium, being a light element, can be easily destroyed in the inner layers of solar-type stars, extending to the outer layers if an efficient mixing process is at work.
The Li abundance is very sensitive to different process such as rotation-induced and overshooting mixing \citep[e.g.][]{Pinsonneault-92, Zhang-12, Xiong-09}.
It also strongly depends on many parameters such as effective temperature, metallicity and age 
\citep[e.g.][]{Pinsonneault-92, DelgadoMena-14, Baumann-10, Carlos-16, Takeda-10, DelgadoMena-15}. The presence of stellar companion can also affect the 
lithium abundance through interactions of the components \citep[e.g.][]{Zahn-94}.
Even the possibility of the Li production by stellar flares have been discussed in the literature \citep{Canal-74, Montes-98}, although the recent observations 
by \citet{Honda-15} does not provide any evidence of Li production by superflares.

Together with the aforementioned processes, it was suggested that the presence of planets and/or formation of planets can also affect the Li content. In particular,
several works, starting from \citet{King-97}, showed that solar analogs (in the temperature range of \teff =\ $T_{\mathrm{\odot}} \pm 80 K$
but a relatively large range of metallicities) with detected planets 
are systematically more depleted in Li than their 'single' counterparts \citep[e.g.][]{Israelian-09, Israelian-04, Chen-06, Gonzalez-10, Gonzalez-08, Gonzalez-15, Takeda-10, 
Castro-09, Figueira-14, DelgadoMena-14}. This relation, however, was contested by several authors \citep[e.g.][]{Ryan-00, Luck-06,  Baumann-10, Ghezzi-10, Ramirez-12, Carlos-16}
arguing that the reported Li depletion in planet hosts relative to the non-hosts can be related to the bias in age, mass and metallicity. 
\citet{Figueira-14}  applied a multivariable regression to simultaneously consider the impact of different parameters (age, metallicity, \teff) on Li abundances.
The authors reached the conclusion that planet-hosting stars display a depletion in lithium.

\begin{figure}
\includegraphics[width=\linewidth]{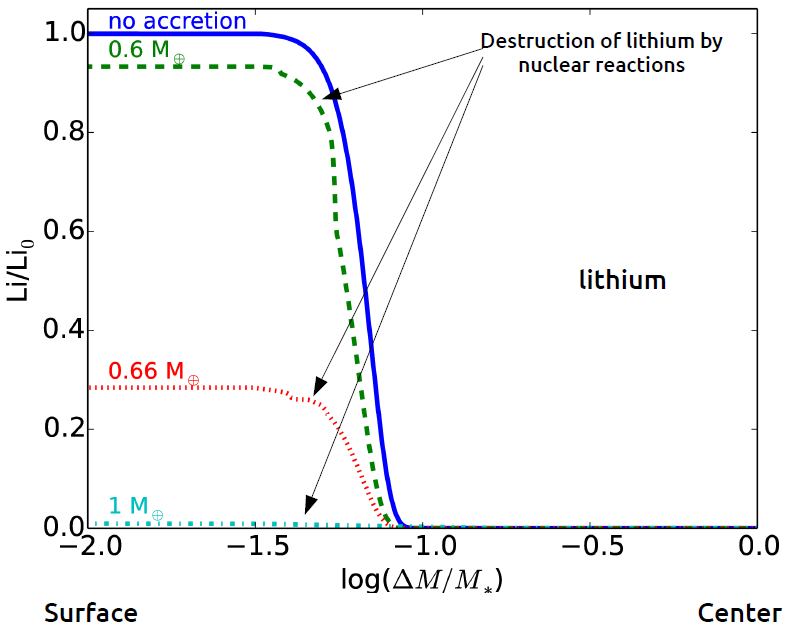}
\caption{Li abundance profiles after the accretion of different
masses at the beginning of the main sequence in the model of 16 Cyg B  (Courtesy of Morgan Deal).
An accreted mass lower than 0.6$\oplus$ practically does not affect the lithium abundance, while 
an accretion of Earth-like chemical composition matter of 0.66$\oplus$ mass is enough to explain the lithium abundance difference observed between the two stars.
}
\label{figure_li}
\end{figure}

As in the case of \tc \ trend, studying stellar twins in binary systems can help to understand the origin of Li depletion. Probably the most suitable system for this 
kind of studies is the 16 Cyg binary system. The 16 Cyg system is composed of two solar-type stars which are two of the best observed \textit{Kepler}
targets. A red dwarf is in orbit around 16 Cyg A, and 16 Cyg B hosts a giant planet. The Li
abundance is much more depleted in 16 Cyg B than in 16 Cyg A, by a factor of at least 4.7 \citep{King-97}. The interesting aspect of studying
the 16 Cyg system is that the two stars have the same age and may be assumed to have the same chemical composition.
Since the observable parameter difference between the two stars is very small (e.g. $\lesssim 0.05 M_{\odot}$ in mass, $\lesssim 80 K$ in temperature), 
the currently observed differences in Li abundance is likely to be 
due to their different evolution, related to the fact that one of them hosts a giant planet while the other does not. 

The fact that 16 Cyg B has a planet suggests that a disk may have been in interaction with the star
at the beginning of its evolution. Recently \citet{Deal-15} studied the impact of the accretion of metal rich planetary
matter onto this star. The accretion modifies the surface chemical composition of the star and may
trigger an instability called fingering (or thermohaline) convection \citep{vauclair04,garaud11,theado12,deal13}. This
instability occurs in the case of a stable temperature gradient and an unstable mean molecular
weight gradient when the thermal diffusivity is larger than the molecular one. This mixing process
dilutes the accreted matter and may transport light elements down to their nuclear destruction layers
and lead to an extra depletion at the surface. The authors used the \cite{brown13} 1D prescription
(determined from 3D simulations) to compute the effect of fingering convection.

\section{Summary}

Determining both precise and accurate stellar abundances is a truly difficult task. There are many different choices to make: 
telescopes and instruments, atomic and molecular data, 1D or 3D model atmospheres that incorporate either LTE or NLTE line 
formation, and techniques that determine abundances with respect to the Sun or in a (line-by-line) differential approach with 
respect to another star. The results from these varying methods produce abundances that can be highly precise, approaching 
$\lesssim$ 0.01 dex, with exciting new findings as discussed above.

\citet{Deal-15} used the TGEC, which includes complete
atomic diffusion (including radiative accelerations). By testing the accretion of planetary matter with
the same chemical composition as the bulk Earth \citep{allegre95}, they found that the more massive the accreted mass, the more  Li depletion occurs at
the surface (see Fig.~\ref{figure_li}) i.e. opposite to a common expectation that the accretion of planetary material 
should increase the Li abundance. The accretion of a fraction of an Earth mass is
enough to explain a Li ratio of 4.7 in the 16 Cyg system.  The authors concluded that such a process may be frequent 
in planet-hosting stars and should be studied in other cases in the future.

It is difficult to determine the highest achievable accuracy in stellar abundance determinations due to the fact that different models and 
analyses are not always in agreement. Additionally, it is complicated to calculate the associated error budget including systematics. 
Going beyond 0.01 dex will likely require modelling of stellar (magnetic) activity \citep[e.g.][]{Fabbian-15} 
in some cases even time-dependent phenomena\footnote{See \citet{Dupree-16} for an extreme case.} like diffusion \citep[e.g.][]{Onehag-14}.
Employing these techniques requires meticulous work and will be limited to relatively small 
data sets in order to enable extremely high accuracy. It is important, for the sake of measuring the true 
surface abundances of stars, that we continue to work on high precision spectroscopy while developing the modelling 
techniques to a higher degree of self-consistency.

Studying stellar abundances allows a deep insight into the formation and evolution of stars and stellar systems. 
Namely, [X/Fe]-age correlations can relate whether stars were formed from well-mixed molecular clouds or 
within areas that were enriched to varying degrees by supernovae \citep{Nissen-16}. By looking at the abundances with age, it is possible 
to get a clearer picture of the nucleosynthetic enrichment timescales for different element classes \citep{Spina-16a}.  
However, studying how stellar abundances vary when a star hosts planets is not straightforward. 
There is an on-going controversy as to whether refractory elements are locked up inside of planets as they form, 
as shown by the \tc \ trend from some studies, but not from others. The solution may be correlated with stellar age, 
Galactocentric distance, or due to a molecular cloud already depleted in refractory elements prior to star formation. 
The Li content, in particular, may be depleted in planetary hosts compared to non-hosts. These questions are 
intriguing because the solutions offer a wide range of stellar and planetary evolution scenarios. 
With a coordination between accurate observed stellar abundances and detailed models for both stars and planets, 
we are optimistic that the mysteries underlying the varying abundance characteristics of Sun-like stars 
unlike the Sun may be revealed in the future.

\acknowledgements
We thank the science organizing committee of Cool Stars 19 for selecting this splinter session and the local organizing committee for the 
provided support in organization of the splinter. We also thank all the 
participants of the splinter session for very active and productive discussion.
V.A. and E.D.M acknowledge the support from Funda\c{c}\~ao para a Ci\^encia e Tecnologia (FCT)
through national funds and from FEDER through COMPETE2020 by the following grants
UID/FIS/04434/2013 \& POCI-01-0145-FEDER-007672, PTDC/FIS-AST/7073/2014 \& POCI-01-0145-FEDER-016880 and PTDC/FIS-AST/1526/2014 \& POCI-01-0145-FEDER-016886.
V.A. also acknowledges the support from FCT through Investigador FCT contracts of reference IF/00650/2015/CP1273/CT0001. E.D.M further acknowledges
the support of the FCT (Portugal) in the form of the grant SFRH/BPD/76606/2011.
NRH acknowledges that the results reported herein benefited from collaborations and/or information exchange within NASA's Nexus for 
Exoplanet System Science (NExSS) research coordination network sponsored by NASA's Science Mission Directorate. In addition, 
N.R.H. also benefited from support of the Vanderbilt Office of the Provost through the Vanderbilt Initiative in Data-intensive Astrophysics (VIDA) fellowship. 
JIGH acknowledges financial support from the Spanish Ministry of Economy and Competitiveness (MINECO) under the 2
013 Ram\'on y Cajal program MINECO RYC-2013-14875, and the Spanish ministry project MINECO AYA2014-56359-P.
Funding for the Stellar Astrophysics Centre is provided by The Danish National Research Foundation (Grant DNRF106).
PGB acknowledges the ANR (Agence Nationale de la Recherche, France) program IDEE (n$^\circ$ ANR-12-BS05-0008) 
"Interaction Des Etoiles et des Exoplanetes" and received funding from the CNES grants at CEA.
  

%
\bibliographystyle{an}
\bibliography{references.bib}


\end{document}